\documentclass[aps,prl,reprint,twocolumn]{revtex4-2}

\newcommand{\ndens}{\SI{2.0e11}{cm^{-2}}}
\newcommand{\gfactor}{3.0}
\newcommand{\tauzero}{\SI{100}{ps}}

\newcommand{\gpint}{\SI{0.2}{\micro eV \micro m^2}}
\newcommand{\gint}{\SI{5}{\micro eV \micro m^2}}

\newcommand{\maxB}{\SI{2.3}{T}}
\newcommand{\Tc}{\SI{2.3}{K}}

\usepackage{graphicx} 
\usepackage{bm}
\usepackage{amsmath}
\graphicspath{{./figures/}}
\usepackage{hyperref}
\hypersetup{colorlinks=true}
\usepackage{titlesec}
\usepackage[capitalise]{cleveref}
\usepackage{amssymb}
\usepackage{siunitx}
\usepackage{newtx}
\usepackage{chemformula}

\usetikzlibrary{external}
\newcommand{\FM}{\mathrm{FM}}

\DeclareMathOperator{\Imm}{Im}

\newcommand{\ii}{\mathrm{i}}
\makeatletter \renewcommand\d[1]{\ensuremath{%
		\mathrm{d}#1\@ifnextchar\d{\!}{}}\;}
\makeatletter \newcommand\D[1]{\ensuremath{%
			\mathcal{D}#1\@ifnextchar\d{\!}{}}\;}
\makeatother
\begin{document}
	\title{Magnetic noise of a dark exciton Bose-Einstein condensate }
	\author{Pieter M. Gunnink}
	\email{pgunnink@uni-mainz.de}

	\affiliation{Institute of Physics, Johannes Gutenberg-University Mainz, Staudingerweg 7, Mainz 55128, Germany}
	
	\date{\today}
	\begin{abstract}
	Excitons provide a promising  platform for the realization of solid-state Bose–Einstein condensation (BEC), offering quantum coherence, strongly correlated electron–hole physics, and superfluidity. Yet, its unambiguous experimental identification remains challenging. 
	In particular, $S_z=\pm1$ triplet excitons are excellent candidates to realize an exciton BEC, because of their intrinsically limited recombination rate and thus long lifetimes. However, since their optical detection is inherently forbidden, experimental signatures remain elusive. In this work, we demonstrate that the magnetic nature of a $S_z=\pm1$ triplet exciton BEC gives rise to stray magnetic field noise, that can be measured using nitrogen-vacancy (NV) center magnetometry. 
	Using an external magnetic field to tune the system from an antiferromagnetic to a ferromagnetic ordering, the longitudinal spin sound mode of the BEC softens---bringing the mode into the characteristic gigahertz frequency range of the NV spin relaxation rate and thus allowing direct detection. Furthermore, we demonstrate an unconventional UV scaling at small distances $d$ from the sample, owing to the cubic corrections to the sound mode dispersion and damping rate in the form of Beliaev damping.  Through this approach, we  establish  that the spin component of exciton BECs offers new approaches to detect exciton BECs.
	 
	\end{abstract}
	\maketitle
	\paragraph{Introduction}
	Bose–Einstein condensation (BEC) is a quantum statistical phase transition, resulting in a macroscopic occupation of the ground state. It was originally observed in liquid helium \cite{allenFlowLiquidHelium1938, kapitzaViscosityLiquidHelium1938} and later in ultracold dilute atomic gases \cite{andersonObservationBoseEinsteinCondensation1995, davisBoseEinsteinCondensationGas1995}, where multicomponent condensates can also be realized \cite{kawaguchiSpinorBoseEinstein2012, stamper-kurnSpinorBoseGases2013}. Excitons provide a promising solid-state platform for BECs \cite{blattBoseEinsteinCondensationExcitons1962,combescotBoseEinsteinCondensation2017}, combining strong interactions, electrical tunability, high temperatures, and multicomponent order, potentially even in equilibrium in excitonic insulators \cite{jeromeExcitonicInsulator1967,kanekoNewEraExcitonic2025}. 
	
	However, unambiguous evidence for a true exciton condensate has not been realized to date, in part due to the recombination processes \cite{butovExcitonCondensationCoupled2003}, which give excitons a finite lifetime. This limits the experimental conditions under which an exciton BEC can be formed. Much of the development in the field of excitonic BECs has therefore been on limiting the recombination rate, such as spatially separating the particle and hole in bilayer heterostructures \cite{foglerHightemperatureSuperfluidityIndirect2014,wuTheoryTwodimensionalSpatially2015,bermanHightemperatureSuperfluidityTwocomponent2016}. 
	\begin{figure}
		\centering
		\includegraphics{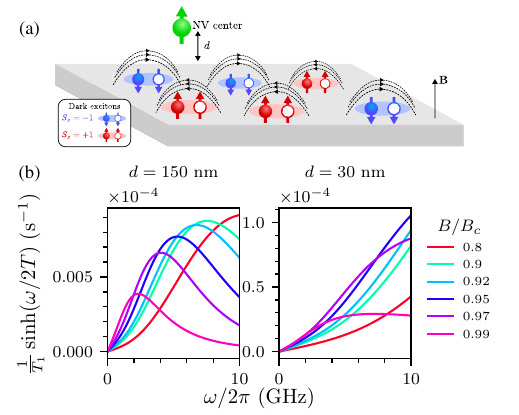}
		\caption{(a) Schematic of magnetic noise of a Bose-Einstein condensate of $S_z=\pm1$ dark excitons, sensed by an NV center at distance $d$ from the two-dimensional sample. An external magnetic field $\bm B$ along $z$ allows for tuning between the antiferromagnetic and ferromagnetic ordering of the exciton condensate.  (b) The reduced transition rate $T_1(\omega)^{-1}\sinh(\omega/2T)$ as a function of frequency, for magnetic fields $B$ approaching the critical field $B_c$, which softens one of the sound modes, bringing it into the measurable frequency range of an NV center ($1$--\SI{10}{GHz}). }
		\label{fig:T1}
	\end{figure}
	
	An alternative approach is to consider optically dark excitons, which do not couple to light and thus have a limited recombination pathway, increasing their lifetime \cite{robertFineStructureLifetime2017}. Focusing on homogeneous BECs, zero-momentum excitons are typically dark if they form a triplet state, for which the electric dipole transition is spin forbidden.
	Two dimensional (2D) transition-metal dichalcogenides (TMDs) are thus excellent candidates for Bose-Einstein condensation of excitons \cite{wangEvidenceHightemperatureExciton2019}: their strong spin-orbit coupling naturally generates dark $S_z=\pm1$ triplet excitons as the lowest energy mode, such as in \ch{W\textit{X}2}, where $\textit{X}$=\ch{S}, \ch{Se} or \ch{Te} \cite{wangColloquiumExcitonsAtomically2018}. In addition, the added spin degree of freedom naturally paves the way for a multicomponent BEC. However, an exciton BEC of optically dark $S_z=\pm1$ triplet excitons  cannot be directly  detected in these compounds via conventional optical means \cite{wangInPlanePropagationLight2017}, and experimental signatures remain elusive. Dark exciton BECs could indeed  already have been formed in existing experiments, but remaining undetected because of the lack of optical signatures \cite{combescotBoseEinsteinCondensationSemiconductors2007,alloingEvidenceBoseEinsteinCondensate2014,anankineQuantizedVorticesFourComponent2017}. 

	In this work, we propose that a BEC of $S_z=\pm1$ triplet excitons has unique magnetic noise signatures associated with its collective sound excitations. As the system is tuned between antiferromagnetic (AFM) and ferromagnetic (FM) excitonic ordering, one of the collective sound modes softens. We propose to use NV center magnetometry to detect this softening, which we demonstrate to be well suited for the energy and length scales relevant to the problem. Finally, we show that the Beliaev damping is enhanced close to the phase transition, and gives rise to an unconventional UV scaling as a function of the distance of the NV center to the sample---providing a potential condensed-matter platform to study fundamental many-body decay processes beyond ultracold atomic gases.

	\paragraph{Method}
	We consider a dilute $S_z=\pm1$ exciton gas, described by the action \footnote{See Supplemental Material at [URL inserted by publisher] for detailed calculation of the BEC, its excitations, Beliaev damping, Landau damping, NV center magnetometry and the low-frequency scaling.}
	\begin{multline}
		S=\int\d{x} \Biggl\lbrace\sum_{m=\pm1}\phi_{m}^\dagger(x)\left[\partial_\tau+\Delta
		-\frac{\nabla^2}{2m} -\mu
		\right]\phi_{m}(x)\\
		-\gamma  B_z(x)  S_z(x)+\frac12 g n(x)^2 + \frac12 g'S_z(x)^2\Biggr\rbrace
	\end{multline}
	where $\phi_{\pm}(x)$ are the exciton fields for $S_z=\pm1$, $n(x)=\sum_m\phi^\dagger_{m}(x)\phi_m(x)$ is the exciton density, $ S_z(x)=\sum_m\bar\phi_{m}(x)\sigma^z_{mm}\phi_m(x)$ is the $z$-component of the exciton spin density. The interactions are described by $g,g'$ \cite{kawaguchiSpinorBoseEinstein2012}.  Here $x\equiv(\tau,\bm x)$, with $\tau,\bm x$ imaginary time and position respectively, $\Delta$ is the exciton gap, $m$ is the exciton mass, which we assume equal for both $S_z=\pm1$, $\gamma$ is the exciton gyromagnetic ratio, and $\mu$ is the exciton chemical potential.

	We focus here on the dynamical longitudinal magnetic susceptibility, $\chi_{zz}(x-x')\equiv\langle \delta M^z(x)\delta M^z(x')\rangle$ \cite{Note1}, which expanded in exciton operators  reads
	\begin{multline}
		\chi_{zz}(\bm q,\ii\Omega_n)
		=\gamma^2\sum_{mn}\sigma^{z}_{mm}\sigma^z_{nn}\int_0^\beta d\tau e^{\ii\Omega_n\tau}\sum_{\bm k \bm k'}\times\\
		 \langle \mathcal T_\tau\bar \phi_{m;\bm k+\bm q}(\tau)\phi_{m;\bm k}(\tau)\bar\phi_{n;\bm k'-\bm q}(0)\phi_{n;\bm k'}(0) \rangle_c \label{eq:chizz-collective},
	\end{multline}
	demonstrating that the magnetic susceptibility is a collective phenomenon and thus able to measure the collective excitations of the BEC. Here $\mathcal T_\tau$ is the imaginary-time-ordering operator.

	We focus on the regime where the excitons form a Bose-Einstein condensate (BEC), formed by external driving, such as optical injection or exciton-polaritons in microcavities \cite{dengExcitonpolaritonBoseEinsteinCondensation2010, kasprzakBoseEinsteinCondensation2006}, or biasing of an electron-hole bilayer \cite{foglerHightemperatureSuperfluidityIndirect2014,wuTheoryTwodimensionalSpatially2015,maStronglyCorrelatedExcitonic2021,qiTwocomponentExcitonCondensates2026}. We will assume (quasi-)equilibrium, where the excitons have thermalized on timescales faster than those considered here for the NV measurement. Alternatively, the condensate can be intrinsic, where the exciton binding energy exceeds the band gap and the semiconductor is therefore an excitonic insulator \cite{jeromeExcitonicInsulator1967,kanekoNewEraExcitonic2025}; the condensate fluctuations are then described by Gaussian	fluctuations around the mean-field solution \cite{adachiGinzburgLandauActionPolarization2023}.
	
	We expand the exciton fields $\phi_m(x)=\phi_{m0}(\bm x)+\phi_m'(\bm x,\tau)$ \cite{stoofUltracoldQuantumFields2009}, and require that $\phi_{m0}(\bm x)=\langle \phi_m(x)\rangle$, such that the terms linear in $\phi_m'(x)$ vanish. This gives the nonlinear differential equations known as the Gross-Pitaevskii equations
	\begin{multline}
		\left[-\frac{\nabla^2}{2m}- \gamma B \sigma^z_{mm}+gn(x)+\sigma^z_{mm}g'S_{z0}(x)\right]\phi^0_m(\bm x)\\ = \mu\phi^0_m(\bm x),
	\end{multline}
	coupling the spin-up and down densities through $g'$. Depending on the external magnetic field,  we have two possible solutions: (i) for $|B|<B_c$, where $ B_c\equiv g'n/\gamma$,  an antiferromagnetic order, where
	\begin{equation}
		\phi_{\pm 1}^0 = \sqrt{n\frac{1\pm f_z}{2}}, 
	\end{equation}
	with $f_z\equiv\gamma B_z /g' n$ the dimensionless spin polarization; (ii) ferromagnetic order, where 
	\begin{equation}
		\phi_{+1}^0 = \sqrt{n};\quad \phi_{-1}^0=0 
	\end{equation}
	and the spin polarization is $f_z=1$. In both phases, the exciton density is given by $n=(\mu-\Delta)/g$.
	
	To provide estimates for these parameters, we follow the recent experimental evidence for two-component exciton condensates in a TMD bilayer \cite{qiTwocomponentExcitonCondensates2026}, thus setting $g=\gint$ and $g'=\gpint$, $m= m_e$ and $\gamma=\gfactor\times \mu_B$ \cite{molasProbingManipulatingValley2019, robertFineStructureLifetime2017, robertMeasurementSpinforbiddenDark2020}, with $\mu_B$ the Bohr magneton, and $n=\ndens$. This gives a critical temperature $T_c\approx 1.3n/m=\Tc$ for a Berezinskii–Kosterlitz–Thouless (BKT) transition \cite{berezinskiiDestructionLongrangeOrder1971, kosterlitzOrderingMetastabilityPhase1973,prokofevCriticalPointWeakly2001}. With these parameters, we obtain that $B_c=\maxB$, demonstrating that with realistic magnetic fields, both antiferromagnetic and ferromagnetic phases can be reached.

	\paragraph{Excitations.}
	The excitations on top of the BEC are described by Bogoliubov theory. In the antiferromagnetic phase, we obtain that the action quadratic in fluctuations is
	\begin{equation}
		S_{\mathrm{Bog}} = \frac12 \sum_k \vec\phi_k^\dagger\begin{pmatrix}
			(-\ii\omega_n+\epsilon_{\bm k})\hat\sigma_0 + \hat h & \hat h\\
			\hat h & (\ii\omega_n+\epsilon_{\bm k})\hat\sigma_0 + \hat h 
		\end{pmatrix}\vec\phi_k
	\end{equation}
	where $\hat h = \frac12 n (g+g') + \bm H\cdot\hat{\bm\sigma}$, with $\bm H=\frac12 n(\sqrt{1-f_z^2}(g-g'),0,f_z(g+g'))$. Here  $\vec\phi(k)=(\phi'_{+1}(k), \phi'_{-1}(k),\bar\phi'_{+1}(-k), \bar\phi'_{-1}(-k))^T$ and hats indicate matrices in the spin space.
	
\begin{figure}
		\centering
		\includegraphics{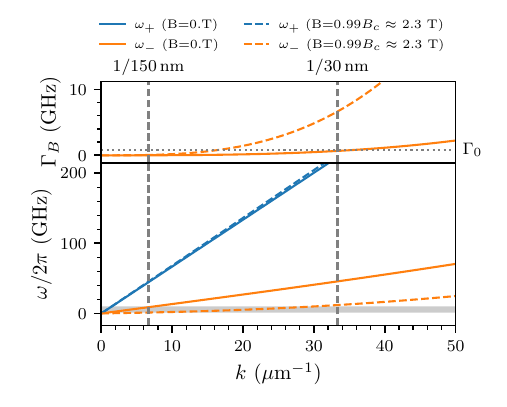}
		\caption{The frequency ($\omega$) of the collective Bogoliubov modes and the Beliaev damping of the lower mode ($\Gamma_B$). The gray shaded area indicates the approximate frequency range over which $T_1$ relaxation noise spectroscopy can be performed. The vertical dashed lines indicate the momenta $k_c\approx 1/d$ which dominate the NV center relaxation noise spectroscopy response, with $d$ the distance between NV center and sample; values are chosen to correspond with \cref{fig:T1}(b). The extrinsic damping $\Gamma_0$ is indicated by the horizontal dotted line, demonstrating that Beliaev damping becomes relevant  in the critical regime . }
		\label{fig:energies}
	\end{figure}

	Performing a Bogoliubov transformation \cite{Note1}, we obtain two modes, with energies
	\begin{equation}
		\omega_\eta = \sqrt{\epsilon_{\bm k}\left(\epsilon_{\bm k}+n(g+g')+\eta n\left[{(g-g')^2+4f_z^2gg'}\right]^{1/2}\right)}.
	\end{equation} 
	For $f_z=0$ these solutions correspond to in- and out-of-phase oscillations of the two spin densities respectively. For finite $f_z$, the two modes are no longer purely in- and out-of-phase oscillations, but they instead mix.  Expanding in small $k$, we obtain that both are gapless sound modes with
	\begin{equation}
		\omega_\eta = c_\eta k + \frac{k^3}{8m^2 c_\eta} + O(k^5),
	\end{equation} 
	and the sound velocity
	\begin{equation}
		c_\eta^2 = {\frac{n(g+g')+\eta\sqrt{n^2(g-g')^2+4f_z^2n^2gg'} }{2m}}.
	\end{equation}
	As the magnetic field approaches the critical value $B_c$, such that $f_z\rightarrow 1$, the lower  sound mode ($\eta=-$) softens and we have that $c_{-1}\rightarrow0$.
	The window in $k$-space where the magnon-like sound mode is linear thus shrinks to zero as the magnetic field approaches the critical magnetic field, such that for $B\approx B_c$ we have that $k^3$ corrections become relevant.
	We show in \cref{fig:energies} the energy spectrum for zero and finite magnetic field, where we also highlight that for finite magnetic field the lower sound mode approaches the gigahertz frequency range which can be measured with NV noise magnetometry.
	
	Finally, we find the retarded longitudinal magnetic susceptibility as \cite{Note1}
	\begin{equation}
		\chi^R_{zz}(\omega,\bm k) = 2\gamma^2 \sum_{\eta} Q_{\eta; k}^2 \frac{\omega_{\eta \bm k}}{(\omega+\ii 0^+)^2-\omega_{\eta \bm k}^2},
	\end{equation}
	where $Q_{\eta k}\equiv e^{-r_{\eta;k}}   \sigma_3\vec\phi_0\cdot\vec\chi^\eta$, with $r_{\eta;k}$ the Bogoliubov angle for mode $\eta$; here $\exp[-r_{\eta;k}]\approx \sqrt{{k}/{2mc_\eta}}$.
	We stress  that the longitudinal magnetic susceptibility has poles at the collective sound mode energies, which is a reflection of the collective nature of  the Bogoliubov quasiparticles. We have disregarded here static contributions to the magnetic susceptibility, i.e., terms that only involve the mean-field solutions $\phi^0_m$, and thermal contributions to the magnetic susceptibility, i.e., the thermal bubble.
%

	In the ferromagnetic phase, we find two modes \cite{Note1}
	\begin{align}
		\omega_{s\FM} &= \sqrt{\epsilon_{\bm k}(\epsilon_{\bm k}+2n(g+g'))}\approx c_{s\FM} k + O(k^3) \\
		\omega_{-1\FM} &= \epsilon_{\bm k}+2\gamma B_z - 2g'n
	\end{align}
	where $c_{s\FM}=\sqrt{\frac{n(g+g')}{m}}$ is the ferromagnetic sound velocity. The first mode is the Bogoliubov sound mode associated with the spontaneous $U(1)$ symmetry breaking of the BEC, whereas the second mode is a massive mode, which vanishes at the critical field $B_c$, connecting to the lower branch in the antiferromagnetic phase. As we demonstrate in the SM \cite{Note1}, the massive mode in the  ferromagnetic phase does not contribute to the $\chi_{zz}$ correlator at zero temperature, whilst the Bogoliubov sound mode has sufficiently large sound velocity such that its signatures can not be detected by $T_1$ relaxation noise spectroscopy. The remainder of this work is thus focused on the antiferromagnetic phase.

	\paragraph{Sensing magnetic susceptibility with an NV center}
	To detect the longitudinal magnetic susceptibility and thus the collective exciton modes, we propose to use here the spin relaxation rate of an NV center \cite{jarmolaTemperatureMagneticFieldDependentLongitudinal2012,rondinMagnetometryNitrogenvacancyDefects2014,casolaProbingCondensedMatter2018,rovnyNanoscaleCovarianceMagnetometry2022,machadoQuantumNoiseSpectroscopy2023} to detect the magnetic noise generated by the exciton BEC \cite{glazovSpinFluctuationsNonequilibrium2016, glazovSpinNoiseExciton2013, ryzhovSpinNoisePolariton2016, smirnovExcitonSpinNoise2014}. Specifically, the Zeeman coupling between the NV center spin and the stray field generated by the exciton BEC induces transitions between the spin states at the resonance frequency $\omega$. The corresponding transition  rate and magnetic field noise are related via a Markovian approximation for the NV dynamics as \cite{flebusQuantumImpurityRelaxometryMagnetization2018,Note1}
	\begin{equation}
					\frac{1}{T_1} = 
		\frac{\gamma_{\rm NV}^2}{8}
		\coth\left[\frac{\omega}{2T}\right]
		\int\frac{d^2k}{(2\pi)^2}
		e^{-2kd}k^2\chi''_{zz}(\omega,\mathbf k), \label{eq:T1}
	\end{equation}
	where $\gamma_{\mathrm{NV}}$ is the NV center gyromagnetic ratio, which we take as $2\mu_B/\hbar$, and $\chi''(\omega,\mathbf k)$ is the negative imaginary part of the retarded longitudinal spin susceptibility. 
	This demonstrates that an NV center can directly probe the magnetic fluctuations generated by the spin-polarized excitons in a semiconductor. We comment here that the $e^{-2kd}k^2$ function has a strong peak at $k=1/d$, and as a first approximation this function can therefore be interpreted as filtering the part of the longitudinal magnetic susceptibility with wavelength $d$. The separation from the sample $d$ can be between \SI{10}{nm} and \SI{1}{\micro \meter} for scanning NV probes, and the $T_1$ relaxation rate can be measured on a gigahertz scale. Furthermore, the NV transition frequency shifts due to the static magnetic field, which could further increase the frequency range over which measurements can be performed \cite{stepanovHighfrequencyHighfieldOptically2015}.

	\paragraph{Beliaev damping}
	Because the linewidth of the excitations directly enters $\Imm[\chi_{zz}]$, we also discuss here the possible sources of damping of the collective excitations. The excitons forming the condensate have an intrinsic lifetime, due to impurity and phonon scattering, as well as radiative decay \cite{moodyIntrinsicHomogeneousLinewidth2015}. Because we consider only $S_z=\pm1$ excitons, which are optically dark, the radiative decay pathway is strongly suppressed \cite{slobodeniukSpinFlipProcesses2016}. This is reflected in experimental observations, where these excitons have significantly lower linewidths compared to their bright counterparts \cite{robertFineStructureLifetime2017,heDispersiveDarkExcitons2025}. Particularly, Ref.~\cite{robertFineStructureLifetime2017} observed a lifetime of \SI{100}{ps} in \ch{WSe2} monolayers for partially dark excitons, indicating that truly dark excitons, whose optical transition is electric dipole forbidden, could have potentially even longer lifetimes.
	
	Given their long lifetimes, we hypothesize here that the intrinsic decay of the Bogoliubov modes could be the major source of damping. At zero temperature, the only allowed decay channel is Beliaev damping \cite{beliaevApplicationMethodsQuantum1957, beliaevEnergySpectrumNonidealBose1957,hodbyExperimentalObservationBeliaev2001, katzBeliaevDampingQuasiparticles2002}. At finite temperature, there exists Landau damping as the result of interactions between the Bogoliubov modes with the non-condensed thermal cloud \cite{popovHydrodynamicHamiltonianNonideal1972,pitaevskiiLandauDampingDilute1997,vincentliuTheoreticalStudyDamping1997}, which we disregard here by focusing on the low-temperature regime. However, we demonstrate \cite{Note1} that Landau damping is expected to give similar results.  Furthermore, we focus on the damping of the $\eta=-$ antiferromagnetic mode, since this is the mode that enters the frequency range of the NV center as the critical magnetic field is approached.
	
	Beliaev damping is associated with the forward self-energy bubble, giving a zero-temperature self energy of
	\begin{equation}
		\Gamma_B(\bm k) = \pi\tilde g^2\int \frac{\d{\bm q}}{(2\pi)^2} |V_{\bm k,\bm q}|^2 \delta(\omega_{\bm k}-\omega_{\bm q}-\omega_{\bm k - \bm q})
	\end{equation}
	where $V_{\bm k,\bm q}$ is the forward-scattering vertex for the scattering of a $\eta=-$ antiferromagnetic mode with momentum $\bm k$ into two $\eta=-$ antiferromagnetic modes with momenta $\bm q$ and $\bm k-\bm q$, and $\tilde g$ is the effective interaction strength \cite{Note1}. We disregard scattering into the $\eta=+$ mode, which is kinematically suppressed due to the strong difference in sound velocities. Performing the integral over $\bm q$ to lowest order in $q$ we obtain  
	\begin{equation}
		\Gamma_B(\bm k) \approx \Gamma_B k^3;\quad \Gamma_B\equiv \sqrt{\frac{3}{2}}\frac{9}{128\pi}\frac{n^{1/4}}{m^{1/2}} \sqrt{\frac{g g'}{g+g'}}\frac{1}{\sqrt{1-f_z}},
	\end{equation}
	where we have expanded in small $1-f_z$ and kept only the lowest order contribution \cite{Note1}. Here, the $k^3$ scaling is a known scaling in two-dimensional BECs \cite{chungDamping2D3D2009}. Importantly, as the magnetic field reaches the critical value $f_z\rightarrow1$, the Beliaev damping diverges and is thus expected to be the dominant contribution to the linewidth. In deriving Beliaev damping it is necessary to keep corrections up to cubic order of the dispersion, and this divergence is therefore not an artifact of the vanishing sound velocity $c_-$. 
	
 	The final damping rate is then $\Gamma=\Gamma_0+\Gamma_B(\bm k)$, where we set $\Gamma_0=\tauzero$ the extrinsic damping lifetime \cite{robertFineStructureLifetime2017}, which is incorporated in \cref{eq:T1} by the substitution $\omega\rightarrow\omega+\ii\Gamma$. We show the strength of the Beliaev damping in \cref{fig:energies}, demonstrating that depending on the momenta probed, the Beliaev damping can exceed the extrinsic damping $\Gamma_0$.
	
	\paragraph{Results}
	We now discuss the resulting transition rates $T_1(\omega)^{-1}$ as measured by an NV center at distance $d$ from the sample, shown in \cref{fig:T1} for magnetic fields approaching the critical magnetic field, and for two distances $d$ from the sample.  
	Focusing first on $d=\SI{150}{nm}$, we observe the typical $T_1(\omega)^{-1}$ response, where the distance $d$ functions as a momentum filter through the $e^{-2kd}$ prefactor in $T_1(\omega)^{-1}$, and is therefore built up mainly from  contributions from $\Imm[\chi_{zz}(1/d,\omega)]$. Increasing the magnetic field towards the critical field, the lower sound mode is lowered in frequency, which brings it into the measurable range of an NV center (1 to \SI{10}{GHz}). Accompanying this lowering of the frequency is also an overall decrease in signal strength, since the $\chi_{zz}$ fluctuations are proportional to $\sqrt{1-f_z}$ \cite{Note1}, and therefore vanish as $f_z$ approaches unity. Furthermore, the linewidth of the collective modes increases due to enhanced Beliaev damping, which is also reflected in the $T_1(\omega)^{-1}$  spectra.

	In realistic setups, it is typically beneficial to be as close as possible to the sample in order to maximize the signal-to-noise ratio. Turning therefore to $d=\SI{30}{nm}$ in \cref{fig:T1}, we observe that even for magnetic fields close to the critical value, the sound mode has frequencies higher than the range accessible by NV center relaxometry. We are therefore only sensitive to the tail of the Lorentzian generated by the sound mode. 
	
\begin{figure}
		\includegraphics{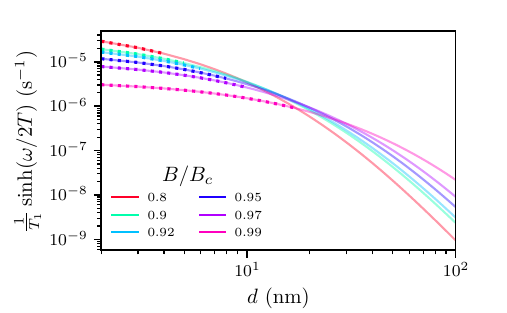}
		\caption{The numerically obtained scaling of the low-frequency ($\omega=\SI{0.1}{GHz}$) reduced transition rate  $T_1^{-1}\sinh(\omega/2T)$ as a function of distance $d$ from the sample, for magnetic fields approaching the critical field. The small $d$-scaling is fitted to $ I_0 + I_1 d + I' d\log d$ (dotted lines), demonstrating the unconventional UV scaling.  }
		\label{fig:d-scaling}
	\end{figure}

	We show in the SM that the scaling of this tail as a function of $d$ contains important information about the $k^3$ scaling of the dispersion and damping of the sound mode. In particular, we show that there is a UV contribution to the small-$d$ scaling, which generates a $\propto d\log d$ scaling. We demonstrate this scaling in \cref{fig:d-scaling}, where we show the low-frequency $T_1(\omega)^{-1}$ transition rate as a function of $d$. The numerically obtained $T_1(\omega)^{-1}$ from \cref{eq:T1} is fitted in the low-$d$ regime, where we define low-$d$ using the crossover momentum where cubic corrections become relevant, $k^*\equiv{c/\alpha}$; here $\alpha\equiv (8m^2c)^{-1}$ is the cubic correction to the dispersion. Explicitly, we show in the SM \cite{Note1} that the scaling in the small-$d$ regime is $\propto I_0 + I_1 d + I'd\log d$.
	Furthermore, we show that scaling of the $d\log d$ contribution is 
	\begin{equation}
		I'\propto \frac{\Gamma_B\alpha}{\alpha^2+\Gamma_B^2},
	\end{equation}
	such that this could be used to extract the ratio  $\alpha/\Gamma_B$ as a function of the magnetic field.
	We demonstrate in  \cref{fig:d-scaling} the accuracy of fitting this scaling for different magnetic fields. We thus conclude that the low-frequency tail contains important information about the exciton condensate, as was also concluded in the context of subgap magnon kinetics and dynamics \cite{duControlLocalMeasurement2017,flebusQuantumImpurityRelaxometryMagnetization2018,fangGeneralizedModelMagnon2022,xueMagnonHydrodynamicsAtomicallythin2025} and more generally for magnetic noise in quantum materials \cite{zhangFlavorsMagneticNoise2022,kellySuperconductivityenhancedMagneticField2024,liNanoscaleObservationControl2026,hammonsProbingGHzSpin2026}.

	\paragraph{Conclusion and discussion.} We have demonstrated that the collective modes of a $S_z=\pm1$ dark exciton Bose-Einstein condensate generate low-frequency magnetic noise, which can be measured using NV magnetometry. Using an external magnetic field, the BEC can be tuned between antiferromagnetic and ferromagnetic ordering. As the critical magnetic field $B_c$ is approached, the softening of the lower collective sound mode generates characteristic noise signatures which can be measured using NV center magnetometry, including an anomalous UV scaling as a function of sampling distance $d$.

	We have considered here only the spin degree of freedom. As experimentally demonstrated in Ref.~\cite{qiTwocomponentExcitonCondensates2026}, considering intervalley exciton interactions gives rise to a richer phase diagram, with intravalley and intervalley condensates, possibly also affecting the magnetic noise of the collective modes. 
	
	An alternative platform is the possible equilibrium exciton condensate in monolayer \ch{WTe2} \cite{jiaEvidenceMonolayerExcitonic2022, sunEvidenceEquilibriumExciton2022}, for which two competing excitons order have been proposed: a spin density wave state and spin spiral order \cite{kwanTheoryCompetingExcitonic2021}. Both would have a characteristic magnetic noise signatures, which could be used to disentangle the two competing orders.
	

	\paragraph{Acknowledgments}
	P.\,M.\,G. is funded by the European Union through an MSCA Postdoctoral Fellowship (Project No. 101145915). P.\,M.\,G. thanks Jairo Sinova and Alexander Mook for valuable feedback on this manuscript.
	\bibliography{noise}
\end{document}


\title{Supplementary material:\\Magnetic noise of a dark exciton Bose-Einstein condensate }
	\author{Pieter M. Gunnink}
	\email{pgunnink@uni-mainz.de}

	\affiliation{Institute of Physics, Johannes Gutenberg-University Mainz, Staudingerweg 7, Mainz 55128, Germany}
	
	\date{\today}
	
	\maketitle
		\section{$S_z=\pm1$ exciton gas}
		We consider a dilute spin-up and spin-down exciton gas, described by the action \cite{stoofUltracoldQuantumFields2009,kawaguchiSpinorBoseEinstein2012}
		\begin{equation}
			S[\bar\phi,\phi]=\int\d{x} \Biggl\lbrace\sum_{m=\pm1}\phi_{m}^\dagger(x)\left[\partial_\tau +\Delta
			-\frac{\nabla^2}{2m} -\mu
			\right]\phi_{m}(x)
			-\gamma  B_z(x)  S_z(x)+\frac12 g n(x)^2 + \frac12 g'S_z(x)^2\Biggr\rbrace \label{eq:action}
		\end{equation}
		where $\phi_{\pm}(x)$ are the exciton fields for $S_z=\pm1$, $n(x)=\sum_m\bar\phi_{m}(x)\phi_m(x)$ is the exciton density, $ S_z(x)=\sum_m\bar\phi_{m}(x)\sigma^z_{mm}\phi_m(x)$ is the $z$-component of the exciton spin density. The interactions are described by $g,g'$.  Here $x\equiv(\tau,\bm x)$ is a shorthand notation, with $\tau,\bm x$ imaginary time and position respectively, $m$ is the exciton mass, which we assume equal for both $S_z=\pm1$, $\gamma$ is the exciton gyromagnetic ratio, $B_z(x)$ is the external magnetic field and $\Delta$ is the exciton gap.
		
		To determine the magnetic susceptibility, we formally introduce the source fields $\bm B_{J}(x)$, such that $\bm B(x)=B_e\hat{\bm z}+\bm B_J(x)$. From these source fields, we can obtain the magnetic susceptibility through functional differentiation as
		\begin{equation}
			\chi_{\alpha\beta}(x,x')\equiv\langle \mathcal T_\tau\delta M^\alpha(x)\delta M^\beta(x')\rangle = \frac{\delta^2\log{Z[B_J]}}{\delta B_J^\beta(x')\delta B_J^\alpha(x)}\bigg |_{B_J=0}; \quad \delta M^\alpha(x)\equiv M^\alpha(x) - \langle M^\alpha(x)\rangle \label{eq:sources} 
		\end{equation}
		where  $\mathcal T_\tau$ is the imaginary-time-ordering operator and
		\begin{equation}
			Z[B_J^\alpha]=\int\mathrm{D}[\bar\phi,\phi]e^{-S[\bar\phi,\phi]}
		\end{equation}
		is the partition function associated with the action \cref{eq:action}. 
		We restrict ourselves to the subspace spanned by the $S_z=\pm1$ excitons, such that $\sigma^{x,y}=0$ and thus only $\langle M^z(x) M^z(x')\rangle \neq0$. We therefore consider here only $\bm B_J(x)=B_J(x)\hat{\bm z}$.	
		Importantly, expanding the longitudinal magnetic susceptibility in the exciton operators we obtain that 
		\begin{equation}
			\chi_{zz}(\bm q,\ii\Omega_n)
			=\gamma^2\sum_{mn}\sigma^{z}_{mm}\sigma^z_{nn}
			\int_0^\beta d\tau e^{\ii\Omega_n\tau}\sum_{\bm k \bm k'} \langle \mathcal T_\tau\bar \phi_{m;\bm k+\bm q}(\tau)\phi_{m;\bm k}(\tau)\bar\phi_{n;\bm k'-\bm q}(0)\phi_{n;\bm k'}(0) \rangle_c
		\end{equation}
		demonstrating that the magnetic susceptibility is a collective phenomenon.

		We expand the exciton fields $\phi_m(x)=\phi_{m0}(\bm x)+\phi_m'(\bm x,\tau)$, and require that $\phi_{m0}(\bm x)=\langle \phi_m(x)\rangle$, such that the terms linear in $\phi_m'(x)$ vanish. This gives the nonlinear differential equations known as the Gross-Pitaevskii equations
		\begin{equation}
			\left[-\frac{\nabla^2}{2m}- \gamma B \sigma^z_{mm}+gn(x)+\sigma^z_{mm}g'S_{z0}(x)\right]\phi^0_m(\bm x) = \mu\phi^0_m(\bm x),
		\end{equation}
		coupling the spin-up and down densities through $g'$. Depending on the external magnetic field,  we have two distinct solutions: (i) for $|B|<B_c$, where $ B_c\equiv g'n/\gamma$ is the critical magnetic field,  an antiferromagnetic (AFM) order, where
		\begin{equation}
			\phi_{\pm 1}^0 = \sqrt{n\frac{1\pm f_z}{2}}, 
		\end{equation}
		with  $f_z\equiv\gamma B_z /g' n$ the dimensionless spin polarization; (ii) ferromagnetic (FM) order, where 
		\begin{equation}
			\phi_{+1}^0 = \sqrt{n};\quad \phi_{-1}^0=0 
		\end{equation}
		and the total magnetization is $f_z=1$.
		Here we have assumed the magnetic field to be positive along the quantization axis. The exciton density is given by $n=(\mu-\Delta)/g$.
		

		\section{Excitations}	
		The excitations on top of the BEC are described by Bogoliubov theory, i.e., by the part of the action that is quadratic in the fluctuations $\phi'_m$. In what follows, we Fourier transform to Matsubara frequencies $\ii\omega_n$ and momenta $\bm k$, using the shorthand $k\equiv(\ii\omega_n,\bm k)$.
		
		\subsection{Antiferromagnetic phase}
		In the antiferromagnetic phase, we obtain that
		\begin{equation}
			S_{\mathrm{Bog}} = \frac12 \sum_k \vec\phi_k^\dagger\begin{pmatrix}
				(-\ii\omega_n+\epsilon_{\bm k})\hat\sigma_0 + \hat h & \hat h\\
				\hat h & (\ii\omega_n+\epsilon_{\bm k})\hat\sigma_0 + \hat h 
			\end{pmatrix}\vec\phi_k
		\end{equation}
		where $\epsilon_{\bm k}=k^2/2m$ is the exciton dispersion, $\hat h = \frac12 n (g+g')\hat\sigma_0 + \bm H\cdot\hat{\bm\sigma}$, with $\bm H=\frac12 n(\sqrt{1-f_z^2}(g-g'),0,f_z(g+g'))$. Here we use the convention that arrows indicate vectors in the full Bogoliubov space, $\vec\phi(k)=(\phi'_{+1}(k), \phi'_{-1}(k),\bar\phi'_{+1}(-k), \bar\phi'_{-1}(-k))^T$, and hats indicate matrices in the spin space. 
		
		We perform a rotation which diagonalizes $\hat h$, i.e., we find
		\begin{equation}
			\hat R^T (\bm H\cdot\hat{\bm\sigma}) \hat R = \hat\sigma_z |\bm H| 
		\end{equation}
		such that 
		\begin{equation}
			S_2 = \frac12\sum_k \tilde\phi_k^\dagger\begin{pmatrix}
				(-\ii\omega_n+\epsilon_{\bm k})\sigma_0 + \tilde h & \tilde h\\
				\tilde h & (\ii\omega_n+\epsilon_{\bm k})\sigma_0 + \tilde h 
			\end{pmatrix}\tilde\phi_k
		\end{equation}
		where $\vec\phi=(\hat\sigma_0\otimes \hat R)\tilde\phi$ and $\tilde h = \hat\sigma_0 \frac12 n (g+g') + \hat\sigma_z |\bm H|$. This amounts to a rotation matrix
		\begin{equation}
			R = [\chi^+ \chi^-];\quad \chi^+=\begin{pmatrix}
				\cos\theta/2 \\ \sin\theta/2
			\end{pmatrix};\quad \chi^-=\begin{pmatrix}
				-\sin\theta/2 \\ \cos\theta/2
			\end{pmatrix}
		\end{equation}
		where $\cos\theta=\frac12 nf_z(g+g')/|\bm H|=\frac{f_z(g+g')}{\sqrt{(g-g')^2+4f_z^2gg'}}$ and $|\bm H|=\frac12 n\sqrt{(g-g')^2+4f_z^2gg'}$. Note that for $f_z$ approaching unity (such that we approach the phase transition between the AFM and the FM phase) we have up to lowest order in $f_z-1$ that
		\begin{align}
			\cos\theta/2 &\approx 1,  \\
			\sin\theta/2 &\approx \frac{g-g'}{\sqrt{2}(g+g')}\sqrt{1-f_z}.
		\end{align}

		Since $\tilde h$ is now diagonal, the two branches decouple and can be diagonalized using the conventional Bogoliubov transformation $\phi_\eta = \cosh r_\eta \beta_\eta - \sinh r_\eta \bar\beta_\eta$ and $\bar\phi_\eta = \cosh r_\eta \bar\beta_\eta - \sinh r_\eta \beta_\eta$, with $\tanh2 r_\eta=(n (g+g')/2+\eta|\bm H|)/(\epsilon_{\bm k}+n(g+g')/2+\eta|\bm H|)$. The corresponding energies are 
		\begin{equation}
			\omega_\eta = \sqrt{\epsilon_{\bm k}(\epsilon_{\bm k}+n(g+g')+n\eta\sqrt{(g-g')^2+4f_z^2gg'})}.
		\end{equation} 
		For $f_z=0$ these solutions correspond to in- and out-of-phase oscillations of the two spin densities respectively. For finite $f_z$, the two modes are no longer purely in- and out-of-phase oscillations, but they instead mix.  Expanding in small $k$, we obtain that these are both gapless sound modes with
		\begin{equation}
			\omega_\eta = c_\eta k + \frac{k^3}{8m^2 c_\eta} + O(k^5).
		\end{equation} 
		and the sound velocity
		\begin{equation}
			c_\eta^2 = {\frac{n(g+g')+\eta\sqrt{n^2(g-g')^2+4f_z^2n^2gg'} }{2m}}
		\end{equation}
		As the magnetic field approaches the critical value $B_c$, such that $f_z\rightarrow 1$, the out-of-phase sound mode velocity reaches zero, indicating the softening of this mode.

		For future reference, we note that
		\begin{equation}
			\cosh r_{\eta k} = \sqrt{\frac{mc_\eta}{2k}} + \frac{1}{2}\sqrt{\frac{k}{2mc_\eta}} + O(k^{3/2});\quad  \sinh r_{\eta k} = \sqrt{\frac{mc_\eta}{2k}} - \frac{1}{2}\sqrt{\frac{k}{2mc_\eta}} + O(k^{3/2}).
		\end{equation}
		
		\subsubsection{Source fields}
		Expanding the source terms, we obtain the contribution to the action
		\begin{equation}
			S_J = -\int\d{x}\gamma B_J^\alpha(x) \sum_{mm'} \sigma_{mm'}^\alpha \left[\phi^0_m \left(\bar\phi'_{m'}(x) + \phi'_{m'}(x)\right) + \phi'_m(x)\bar\phi'_{m'}(x) \right],
		\end{equation}
		and we thus obtain parts of the action that are linear or quadratic in fluctuations. The terms quadratic in fluctuations will describe the thermal bubble, i.e., give rise to terms that are proportional to
		\begin{equation}
			\chi_{\alpha\beta}(\omega,\bm q) \sim \sum_{\bm k} \frac{n_B(\epsilon_{ k})-n_B(\epsilon_{\bm k + \bm q})}{\omega + \ii 0^+ + \epsilon_{\bm k}-\epsilon_{\bm k + \bm q}} \label{eq:bubble}.
		\end{equation}
		Since we focus on the low-temperature regime, we expect that contribution will vanish and we therefore neglect the quadratic in fluctuations terms. 
		
		Applying the rotation as before, the source fields now rotate as
		\begin{equation}
			S_J = -\sum_k\gamma B_{J;-k}^z\sum_\eta\left[ e^{-r_\eta} (\sigma_3\vec\phi_0\cdot\vec\chi^\eta) ({\beta}_{\eta;k}+\bar{\beta}_{\eta;-k}) \right],
		\end{equation} 
		where $\vec\chi^\eta$ are the eigenvectors of $\bm H\cdot \hat{\bm\sigma}$, i.e., columns of the rotation matrix $R=[\vec\chi^+,\vec\chi^-]$. Here we have only kept $B_J^z$.
		The action thus becomes in the rotated and diagonalized frame
		
		\begin{equation}
			S=\sum_{k,\eta}\Big[
			\bar\beta_{\eta k}(-i\omega_n+\omega_{\eta\mathbf k})\beta_{\eta k}
			-\gamma B^z_{J;-k}Q_{\eta k}
			(\beta_{\eta;k}+\bar\beta_{\eta;-k})
			\Big].
		\end{equation}
		where $Q_{\eta k}\equiv  e^{- r _{\eta;k}}  (\sigma_3\vec\phi_0\cdot\vec\chi^\eta)$. We can now integrate out the Bogoliubov modes, and find
		\begin{equation}
			\log\{Z[J]\} = -\gamma^2 \sum_{\eta k}  B_{J;-k} Q_{\eta; -k} \frac{\omega_{\eta \bm k}}{(\ii\omega_n)^2-\omega_{\eta \bm k}^2} B_{J;k} Q_{\eta; k}-\sum_{\eta k}\log(-\ii\omega_n +  \omega_{\eta \bm k} )
		\end{equation}
		and we can directly obtain 
		\begin{equation}
			\langle \delta M^z_{-k} \delta M^z_k\rangle = 2\gamma^2 \sum_{\eta} Q_{\eta; k}^2 \frac{\omega_{\eta \bm k}}{(\ii\omega_n)^2+\omega_{\eta \bm k}^2},
		\end{equation}
		since $Q_{\eta;k}=Q_{\eta;-k}$.  Note that 
		\begin{equation}
			e^{- r _{\eta k}}=\cosh r _{\eta;k}-\sinh r _{\eta;k} =\sqrt{\frac{k}{2mc_\eta}} + O(k^{3/2}).
		\end{equation}
		
		Finally we comment here that as $f_z$ approaches unity, we have that the lower branch contribution the longitudinal magnetic susceptibility vanishes, since
		\begin{equation}
			\sigma_3\vec\phi_0\cdot\vec\chi^- \approx \frac{g-g'}{\sqrt{2}(g+g')}\sqrt{1-f_z} \psi^0_{+1} - \psi^0_{-1} = \sqrt{\frac{n}{2}}\sqrt{1-f_z} \left[\frac{g-g'}{\sqrt{2}(g+g')}\sqrt{1+f_z} - 1
			\right].
		\end{equation}

		\subsection{Ferromagnetic phase}
		In the ferromagnetic phase, we have that 
		\begin{equation}
			S_{\mathrm{Bog}}=\frac{1}{2}\int\d{k}{\vec\phi_k}^\dagger  \left[
			-\ii\omega_n\sigma_z+\epsilon_{\bm k}\sigma_0  +
			\begin{pmatrix}
				n(g +g') & 0 & n(g +g')  & 0 \\
				0 & -2ng'+2\gamma B_z & 0 & 0 \\
				n(g +g') & 0 & n(g +g') & 0 \\
				0 & 0 & 0 & -2ng'+2\gamma B_z
			\end{pmatrix}
			\right]\vec \phi_k.
		\end{equation}
		We can thus directly read off the Bogoliubov modes as 
		\begin{align}
			\omega_{s\FM} &= \sqrt{\epsilon_{\bm k}(\epsilon_{\bm k}+2n(g+g'))}\approx c_{s\FM} k + O(k^3) \\
			\omega_{-1\FM} &= \epsilon_{\bm k}+2\gamma B_z - 2g'n
		\end{align}
		where $c_{s\FM}=\sqrt{\frac{n(g+g')}{m}}$ is the ferromagnetic sound velocity. The first mode is the Bogoliubov sound mode associated with the spontaneous $U(1)$ symmetry breaking of the BEC, whereas the second mode is a massive mode, which vanishes at the critical field $B_c$, connecting to the lower sound mode in the antiferromagnetic phase.
		
		The source fields are given by 
		\begin{equation}
		S_J=-\gamma\sum_{k,\eta}B^z_{J;-k}
		Q^z_{\mathrm{FM};\eta k}
		(\beta_{\eta;k}+\bar\beta_{\eta;-k}).
		\end{equation}
		where 
		\begin{equation}
			Q^z_{\FM;\eta}=\begin{cases}
				-\sqrt{{n\epsilon_{\bm k}}/{\omega_{s; \bm k}}};\quad &\eta=s\\
				0;\quad & \eta={-1}
			\end{cases}
		\end{equation}
		We thus obtain that only the Bogoliubov sound mode ($\eta=s$) contributes to the longitudinal magnetic susceptibility
		\begin{equation}
			\langle \delta M^z_{-k} \delta M^z_k\rangle_{\mathrm{FM}} = 2\gamma^2 Q_{s; k}^2 \frac{\omega_{s; \bm k}}{(\ii\omega_n)^2+\omega_{s; \bm k}^2},
		\end{equation}
		since $Q_{-1; k}=0$. This can be traced back to the fact that magnetic susceptibility is a collective effect, consisting of spin fluctuations in combination with the condensate, but we have in the FM phase that $\psi_{-1}^0=0$. The only remaining contribution is therefore from the $\omega_{s\FM}$ mode, which has a sufficiently large sound velocity such that its signatures cannot be detected by $T_1$ relaxation noise spectroscopy. We therefore focus in this work on the antiferromagnetic phase. We finally note that at finite temperature, there will be signatures of the $\eta=-1$ mode in the longitudinal magnetic susceptibility, due to the thermal bubble [\cref{eq:bubble}], but again we expect these to be typically small.
	
		\section{Beliaev damping}
		At zero temperature, the only source of damping within the condensate is due to Beliaev damping \cite{beliaevApplicationMethodsQuantum1957, beliaevEnergySpectrumNonidealBose1957}, which is associated with the decay of one Bogoliubov mode into two Bogoliubov modes. We therefore expand the interaction up to third order to find
		\begin{equation}
			S_3 = \sqrt{n}\int\d{x}\sum_{\gamma=\pm }\left(gn'(x)+\gamma g'S_z'(x)\right)\sqrt{1+\gamma f_z}(\phi_{\gamma}'(x)+\bar\phi_{\gamma}'(x)) 
		\end{equation}
		We now rotate using the matrix $R$ again, to find
		\begin{equation}
			S_3 = \sqrt{n}\int\d{x}\sum_{\gamma\eta}\left(g\tilde n(x)+\gamma g'\tilde S_z(x)\right)\sqrt{1+\gamma f_z} \chi^\eta_{\gamma}(x)(\tilde \phi_\eta(x) + \hc)
		\end{equation}
		where $\tilde n =  \tilde \bar\phi(x) \tilde \phi(x)$ and $\tilde S_z=\tilde \bar\phi(x) (R^T\sigma_z R)\tilde \phi(x)=\cos\theta  \bar{\tilde\phi}\sigma_z\tilde\phi-\sin\theta \bar\phi\sigma_x\tilde\phi$. We focus on Beliaev damping of the critical branch ($\eta=-$) as $f_z$ approaches 1, so only scattering within the critical branch is kinematically allowed. We thus keep only $\tilde n \approx \bar{\tilde\psi}_-\tilde \psi_-$ and $\tilde S_z \approx -\cos\theta \bar{\tilde\psi}_-\tilde \psi_- $,\footnote{In what follows, we drop the $\eta$ labels and assume all operators to be $\eta=-$.} and obtain
		\begin{equation}
			S_3 = \sqrt{n}\int\d{x}\sum_{\gamma} \left(g - \gamma g'\cos\theta \right)\sqrt{1+\gamma f_z} \chi^-_{\gamma} \tilde n (x)(\tilde \psi (x) + \hc)
		\end{equation}
		which we rewrite as
		\begin{equation}
			S_3 = \tilde g \int\d{x} \tilde n_- (x)(\tilde \psi_- (x) + \hc)
		\end{equation}
		where 
		\begin{equation}
			\tilde g=\sqrt{n}\sum_{\gamma} \left(g - \gamma g'\cos\theta \right)\sqrt{1+\gamma f_z} \chi^-_{\gamma}=4\sqrt{n}\frac{g g'}{g+g'}\sqrt{1-f_z} + O(1-f_z)
		\end{equation}
		is the effective interaction vertex, which we have expanded up to first order in $1-f_z$, anticipating the Beliaev damping to be most important near this critical point.

		We now Fourier transform to $k\equiv(\ii\omega_n,\bm k)$,\footnote{Where no confusion can arise, we also use the notation $k\equiv|\bm k|$ for the absolute momentum.} and transform to the Bogoliubov modes to find
		\begin{equation}
			S_3 =\frac{\tilde g}{\sqrt{N}} \sum_{kq} V_{\bm k,\bm q}\bar\beta^-_{k-q}\bar\beta^-_{q}\beta^-_{k} + \hc
		\end{equation}
		with the vertex
		\begin{equation}
			V_{\bm k,\bm q}=u^-_{\bm q}\left(
			u^-_{\bm k}u^-_{\bm k - \bm q} + v^-_{\bm k}u^-_{\bm k - \bm q} + u^-_{\bm k}v^-_{\bm k - \bm q}
			\right)+v^-_{\bm q}\left(
			v^-_{\bm k}v^-_{\bm k - \bm q} + v^-_{\bm k}u^-_{\bm k - \bm q} + u^-_{\bm k}v^-_{\bm k - \bm q}
			\right)
		\end{equation}
		where $u_{\bm k}\equiv\cosh r_{\bm k}$ and $v_{\bm k}\equiv-\sinh r_{\bm k}$. We now expand the vertex in the long wavelength limit, and make use of the fact that by energy conservation $k=q+|\bm k - \bm q| + O(k^3)$. Then we have that 
		\begin{equation}
			V_{\bm k,\bm q}=\frac{9}{8}\frac{\sqrt{kq|\bm q - \bm k|}}{(2m\sqrt{n}((g+g')-\sqrt{(g-g')^2+4f_z^2gg'}))^{3/4}}
		\end{equation}
		which coincides with the canonical result \cite{giorginiDampingDiluteBose1998,chungDamping2D3D2009}.
		
		\begin{figure}
			\centering
			\includegraphics[width=0.25\textheight]{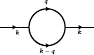}
			\caption{The forward self-energy, associated with Beliaev damping.}
			\label{fig:circle}
		\end{figure}
		
		The Beliaev damping then follows from the imaginary part of the forward self-energy bubble, as shown in \cref{fig:circle}, which is
		\begin{equation}
			\Imm[\Sigma_f(\bm k)] = \pi\tilde g^2\int \frac{\d{\bm q}}{(2\pi)^2} |V_{\bm k,\bm q}|^2 \delta(\omega_{\bm k}-\omega_{\bm q}-\omega_{\bm k - \bm q}) \approx \pi\tilde g^2 |V|^2 \int\frac{\d{\bm q}}{(2\pi)^2} kq|\bm q - \bm k| \delta(\omega_{\bm k}-\omega_{\bm q}-\omega_{\bm k - \bm q})
		\end{equation}
		with $V=\frac{9}{8}\frac{1}{(4m^2n)^{3/8}((g+g')-\sqrt{(g-g')^2+4f_z^2gg'})^{3/4}}$, which is in the limit of $f_z\rightarrow1$ gives
		\begin{equation}
			V^2\approx \frac{81}{1024\sqrt{2}}\left(\frac{g+g'}{\sqrt{n}mg g'(1-f_z)}\right)^{3/2} + O((1-f_z)^{-1/2})
		\end{equation}
		
		To evaluate the integral 
		\begin{equation}
			I=\int \d{\bm q} kq|\bm q - \bm k| \delta(\omega_{\bm k}-\omega_{\bm q}-\omega_{\bm k - \bm q}),
		\end{equation}
		we change variables to $p=\sqrt{k^2 + q^2 - 2kq\cos\theta}$, such that 
		\begin{equation}
			I=2\int \d{q}\int_{k-q}^{k+q}\d{p} \frac{p^2 q}{\sqrt{1-(k^2+q^2-p^2)^2/(2kq)^2}} \delta(\omega_{k}-\omega_{q}-\omega_p) 
		\end{equation} 
		where we have additionally reduced the integral over $\theta=[0,2\pi]$ to $\theta=[0,\pi]$, and made use of the fact that $q<k$.
		
		The zeros of $\omega_{k}-\omega_{q}-\omega_p$ are,  
		\begin{equation}
			p_*=\sqrt{2m}\left[{-mc^2+\sqrt{(\omega_{k}-\omega_{ q})^2+(mc^2)^2}}\right]^{1/2} \approx (k-q)\left(1+\frac{3qk}{8m^2c^2}\right)
		\end{equation}

		and \begin{equation}
			\frac{\partial\omega_p}{\partial p}\bigr|_{p=p_*}=\frac{p_*(p_*^2/2m^2+c^2)}{\omega_{p_*}} \approx c 
		\end{equation}
		as well as 
		\begin{equation}
			\sqrt{1-(k^2+q^2-p_*^2)^2/(2kq)^2} \approx \frac{\sqrt{3}}{2mc}(k-q)
		\end{equation}
		and therefore we have that
		\begin{equation}
			I=2\int_0^k \d{q}\left(\frac{\partial\omega_p}{\partial p}\bigr|_{p=p_*}\right)^{-1}\frac{p_*^2 k }{\sqrt{1-(k^2+q^2-p_*^2)^2/(2kq)^2}} \approx \frac{4m}{\sqrt{3}}\int_0^k \d{q}(k-q)q
		\end{equation}
		so we have
		\begin{equation}
			I =  \frac{2m}{3\sqrt{3}}k^3,
		\end{equation}
		as was also shown by Refs.~\cite{chungDynamicalStructureFactor2008,chungDamping2D3D2009}.

		Finally, we thus obtain that 
		\begin{equation}
			\Imm[\Sigma_f(\bm k)] =  \frac{m \tilde g^2 V^2}{6\pi\sqrt{3}}k^3\approx\sqrt{\frac{3}{2}}\frac{9}{128\pi}\frac{n^{1/4}}{m^{1/2}} \sqrt{\frac{g g'}{g+g'}}\frac{k^3}{\sqrt{1-f_z}}
		\end{equation}
		demonstrating that the Beliaev damping for the $\omega_-$ branch diverges as the critical field is approached---as is to be expected, since Beliaev damping depends on the finite $k^3$ curvature of the dispersion, which diverges in the limit of $f_z\rightarrow 1$. Note that our calculation is not valid at $f_z=1$. For notational simplicity, we write 
		\begin{equation}
			\Imm[\Sigma_f(\bm k)] = \Gamma_B k^3;
			\quad  \Gamma_B\equiv \sqrt{\frac{3}{2}}\frac{9}{128\pi}\frac{n^{1/4}}{m^{1/2}} \sqrt{\frac{g g'}{g+g'}}\frac{1}{\sqrt{1-f_z}}.
		\end{equation}
		
		\section{Landau damping}
		Landau damping is described by the thermal circle diagram, where a Bogoliubov mode scatters from a thermally populated mode into another mode. We focus on the Landau damping process within the lower branch $\eta=-1$, but we note here that kinematics also allows for the process involving an internal thermal $\eta=-1$ and forward-propagating $\eta=+1$ mode.
		
		Then we have that 
		\begin{equation}
			S_3 =\frac{\tilde g}{\sqrt{N}} \sum_{kq} V_{\bm k,\bm q}\bar\beta^-_{q}\beta^-_{k+q}\beta^-_{k} + \hc
		\end{equation}
		with the Bogoliubov vertex
		\begin{equation}
			V_{\bm k,\bm q}=u^-_{\bm q}\left(
			u^-_{\bm k}u^-_{\bm k + \bm q} + v^-_{\bm k}v^-_{\bm k + \bm q} + u^-_{\bm k}v^-_{\bm k + \bm q}
			\right)+v^-_{\bm q}\left(
			v^-_{\bm k}v^-_{\bm k + \bm q} + u^-_{\bm k}u^-_{\bm k + \bm q} + v^-_{\bm k}u^-_{\bm k + \bm q}
			\right).
		\end{equation}
		Importantly, the effective vertex $\tilde g$ also enters the Landau damping, which goes as $\sqrt{1-f_z}$. Since we only consider scattering within the $\eta=-1$ branch, and our interaction vertex $\tilde g$ is momentum-independent, we can use the results from Ref.~\cite{chungDamping2D3D2009} to find in the low temperature limit ($T\ll mc^2$) that 
		\begin{equation}
			\Gamma_L = \frac{\sqrt{3}\pi }{4} \frac{g+g'}{ngg'}\frac{1}{1-f_z}( T)^2\omega_{\bm k}
		\end{equation}
		demonstrating that the Landau damping also diverges as $f_z\rightarrow1$, whilst its momentum scaling is determined by the sound mode dispersion. 
		\section{NV center}
		To detect the magnetic susceptibility, we propose to use here the spin relaxation rate of an NV center. Specifically, we have that, making use of the fluctuation dissipation theorem
		\begin{equation}
			\frac{1}{T_1} = \frac{\gamma_\mathrm{NV}^2}{2}\coth\left[\frac{\omega}{2T}\right]\chi_{B^-, B^+}''(\omega), 
		\end{equation}
		where 
		\begin{equation}
			\chi_{B^-, B^+}''(\omega) = -\Im \chi^R_{B^-, B^+}(\omega)
		\end{equation}
		is the negative imaginary part of the retarded susceptibility,
		\begin{equation}
			\chi^R_{B^-, B^+}(\omega)=-\ii\int_0^\infty \d{t} e^{\ii\omega t} \langle [\hat B^-(t),\hat B^+(0)]\rangle,
		\end{equation}
		where $\hat B^\pm(t)$ are the magnetic field operators at the position of the NV center and $\gamma_\mathrm{NV}$ is the gyromagnetic ratio of the NV center, which we take to be $\gamma_\mathrm{NV}=2\mu_B/\hbar$.
		This can be directly obtained from the magnetization in the semiconductor as 
		\begin{equation}
			\hat{\bm B}(\bm r) = -\frac{\mu_0}{4\pi} \left[\frac{\hat{\bm m}}{r^3}-3\frac{(\bm r\cdot\hat{\bm m})\bm r}{r^5}\right]
		\end{equation}
		Expressing the magnetization operators in momentum space, $\hat{\bm m}(\bm r)=\frac{1}{\sqrt{N}} \sum_{\bm k}e^{\ii \bm k \cdot \bm r}\hat{\bm m}_{\bm k}$, we obtain that 
		\begin{equation}
			\hat B^\pm(\mathbf r_{\rm NV})=
			\int\frac{d^2k}{(2\pi)^2}
			\left(f_{\mathbf k}^x\pm i f_{\mathbf k}^y\right)
			\hat M^z_{\mathbf k}.
		\end{equation}
		where $f_{\bm k}^\eta = \frac{3d}{4\pi} \sum_{\bm r} e^{\ii\bm k\cdot \bm r} r_\eta/r^5 =\frac\ii2 e^{-|\bm k|d} k_\eta $ for $\eta\in\{x,y\}$. Here we have assumed that the NV axis to be parallel to the spin quantization axis of the $S_z=\pm1$ excitons. We can now directly relate the $T_1$ relaxation rate to the longitudinal magnetic susceptibility as
		\begin{equation}
			\frac{1}{T_1} = 
			\frac{\gamma_{\rm NV}^2}{8}
			\coth\left[\frac{\omega}{2T}\right]
			\int\frac{d^2k}{(2\pi)^2}
			e^{-2kd}k^2\chi''_{zz}(\omega,\mathbf k)
		\end{equation}
		This demonstrates that an NV center can directly probe the magnetic fluctuations generated by the spin-polarized excitons in a semiconductor. We comment here that the $e^{-2qd}q^2$ function has a strong peak at $q=1/d$, and as a first approximation this function can therefore be interpreted as filtering the part of the longitudinal magnetic susceptibility with wavelength $d$. 
		
		\section{Low-frequency noise enhancement as $f_z\rightarrow1$}
		As $f_z$ approaches unity, the sound velocity of the lower AFM branch vanishes, and therefore the cubic corrections to the dispersion become relevant. We demonstrate here that this leads to a strong low-frequency contribution. 

		For finite damping, we have that  $\Imm[\chi_{zz}(\omega,\bm q)]=  4\omega \omega_q \Gamma_q / [(\omega^2-\omega_q^2-\Gamma_q^2)^2+4\omega^2 \Gamma_q^2]$. We then have that in the limit of $\omega\rightarrow0$
		\begin{equation}
			\lim_{\omega\rightarrow0}\frac{1}{T_1} \sim\frac{ T}{4\pi mc}\int  \d{q} e^{-2qd}q^4 \frac{4\omega_q \Gamma_q}{(\omega_q^2+\Gamma_q^2)^2},
		\end{equation}
		where $\Gamma_q = \Gamma_0 + \Gamma_Bq^3$.
	There are two natural momentum scales, set by $k^*_{\alpha}\equiv \sqrt{c/\alpha}$ and $k^*_{\Gamma}\equiv (\Gamma_0/\Gamma_B)^{1/3}$, which are the momentum scales where the dispersive and dissipative cubic corrections dominate. Here $\alpha\equiv (8m^2c)^{-1}$ is the cubic correction to the dispersion. 
	We focus on the low damping limit, such that $\Gamma_0$ is small and thus $k^*_{\Gamma}\ll k^*_{\alpha}$, such that we have $k^*\equiv\max[k^*_{\Gamma},k^*_{\alpha}]=k^*_{\alpha}$ as the momentum scale where the cubic corrections are dominant.  In the limit of $1/d\gg k^*$ we then have that the integral
	\begin{equation}
		I(d) \equiv \int_{0}^\infty \d{q} e^{-2qd}q^4\frac{4\omega_q \Gamma_q}{(\omega_q^2+\Gamma_q^2)^2},
	\end{equation}
	can be rewritten as
	\begin{equation}
		I(d) = I_0 - I'(d);\quad I_0\equiv \int_{0}^\infty \d{q}q^4\frac{4\omega_q \Gamma_q}{(\omega_q^2+\Gamma_q^2)^2};\quad I'_d\equiv \int_{0}^\infty \d{q} (1-e^{-2qd})q^4\frac{4\omega_q \Gamma_q}{(\omega_q^2+\Gamma_q^2)^2}.
	\end{equation}
	Here $I_0$ will give an independent of $d$ contribution. The $I'_d$ term can be expanded further to give 
	\begin{equation}
		I'_d \approx 2d\int_{0}^\infty \d{q} q^5\frac{4\omega_q \Gamma_q}{(\omega_q^2+\Gamma_q^2)^2}.
	\end{equation}
	Now there exist two contributions to this integral: the low-$q$ and high-$q$ regions. We split this therefore further in $I_d'=I_d^{\mathrm{IR}} + I_d^{\mathrm{UV}}$. For low-$q$, we can expand the integral in small $q$ and directly find the IR contribution as
	\begin{equation}
		I_d^{\mathrm{IR}} \approx 2d\int_{0}^{k_*} \d{q} \frac{4c}{\Gamma_0^3}q^6 =\frac{8c}{7\Gamma_0^3}\left(k_*\right)^7 d 
	\end{equation}
	to give a linear in $d$ contribution.
	
	For the high-$q$ tail, we note that the main contributions come from large $q$, and we therefore keep only the large $q$ terms in the integral to find the UV contribution
	\begin{equation}
		I_d^{\mathrm{UV}} = 8d\frac{\Gamma_B\alpha}{\alpha^2+\Gamma_B^2}\int_{k_*}^{1/d} \d{q} \frac{1}{q} =  8d\frac{\Gamma_B\alpha}{\alpha^2+\Gamma_B^2}(\log \frac1d - \log k_*) = 8d\frac{\Gamma_B\alpha}{\alpha^2+\Gamma_B^2}\log \frac{1}{dk_*}
	\end{equation}
	we therefore conclude that in this limit we have that 
	\begin{equation}
		\lim_{\omega\rightarrow0}\frac{1}{T_1} \sim I_0 + I_1 d + 8\frac{\Gamma_B\alpha}{\alpha^2+\Gamma_B^2}d\log d. \label{eq:low-d}
	\end{equation}
	In the main text, we numerically calculate $T_1^{-1}$, and fit \cref{eq:low-d} to the small-$d$ region.
		
	\bibliography{noise}